\documentclass[11pt]{article}

\usepackage{graphicx}
\usepackage{multicol,multirow}
\usepackage{amsmath,amssymb,amsfonts}
\usepackage{mathrsfs}
\usepackage{amsthm}
\usepackage{apacite}
\usepackage{rotating}
\usepackage{appendix}
\usepackage{url}
\usepackage[normalem]{ulem}
\usepackage[authoryear]{natbib}
\usepackage{caption}
\usepackage{subcaption}
\usepackage{float}

\newcommand{\footremember}[2]{%
    \footnote{#2}
    \newcounter{#1}
    \setcounter{#1}{\value{footnote}}%
}
\newcommand{\footrecall}[1]{%
    \footnotemark[\value{#1}]%
} 

\usepackage{geometry}
\begin{document}

\linespread{1.3}

\title{On the Use of Data from Multiple Mobile Network Operators in Europe to fight COVID-19}

\author{Michele Vespe\footnote{Corresponding author, Email: {\tt michele.vespe@ec.europa.eu}} \footremember{trailer}{European Commission,
              Joint Research Centre,
              Via Enrico Fermi 2749, 21027 Ispra (VA), Italy} \and
        Stefano Maria Iacus\footrecall{trailer} \and Carlos Santamaria\footrecall{trailer} \and
        Francesco Sermi\footrecall{trailer} \and
        Spyridon Spyratos\footrecall{trailer} 
}

\date{}

\maketitle

\abstract{The rapid spread of COVID-19 infections on a global level has highlighted the need for accurate, transparent and timely information regarding collective mobility patterns to inform de-escalation strategies as well as to provide forecasting capacity for re-escalation policies aiming at addressing further waves of the virus. Such information can be extracted using aggregate anonymised data from innovative sources such as mobile positioning data. This paper presents lessons learnt and results of a unique Business-to-Government (B2G) initiative between several Mobile Network Operators in Europe and the European Commission. Mobile positioning data have supported policy makers and practitioners with evidence and data-driven knowledge to understand and predict the spread of the disease, the effectiveness of the containment measures, their socio-economic impacts while feeding scenarios at EU scale and in a comparable way across countries. The challenges of this data sharing initiative are not limited to data quality, harmonisation, and comparability across countries, however important they are. Equally essential aspects that need to be addressed from the onset are related to data privacy, security, fundamental rights and commercial sensitivity.}


\paragraph{Policy Significance}
The use of insights derived from Mobile Network Operators revealed the potential of data science and innovation in supporting evidence-driven crisis response and adoption of targeted policy measures to fight the COVID-19 pandemic. This initiative shows that when the use of privately-held data takes place responsibly and is aimed at effectively responding to pressing policy questions, it has the capacity to impact the policy cycle. Even though trust between the involved stakeholders is essential for dealing with this kind of sensitive data, responsible research requires additional verification that, in fact, no harm is created, neither to the assets of the data providers, nor to the legitimate interests of the data subjects involved.

\section{Introduction}\label{Intro}

The new coronavirus disease 2019 (COVID-19) rapidly spread throughout the world during the first quarter of 2020, reaching pandemic status on 11 March 2020. Authorities in most of the European countries and worldwide had to confront with unprecedented challenges to contain the number of infections and prevent saturation of intensive care units in national health systems. This required immediate policy responses. Governments reacted by passing a wide range of measures, including confinement measures designed to contain the spread of the virus, information campaigns, fiscal stimulus to support the economy in the short term, recovery plans for the aftermath and preparation to prospective second waves of the virus. The need for timely, accurate and reliable data that would inform such decisions is of paramount importance.

Against this backdrop, on 8 April the European Commission asked European Mobile Network Operators (MNOs) to share anonymised and aggregate mobile positioning data. The aim was to provide mobility patterns of population groups and serve the following purposes in the fight against COVID-19. Initiated by means of an exchange of letters, the terms of cooperation between MNOs and the European Commission are outlined by a Letter of Intent\footnote{European Commission and GSMA partners on Data4Covid: \url{https://www.gsma.com/gsmaeurope/resources/d4c/}}, which specifies that insights into mobility patterns of population groups extracted in the framework of this initiative are meant to serve the following purposes:

\begin{itemize}
    \item ``understand the spatial dynamics of the epidemics thanks to historical matrices of mobility national and international flows;
    \item quantify the impact of social distancing measures (travel limitations, non-essential activities closures, total lock-down,etc.) on mobility;
    \item feed SIR epidemiological models, contributing to the evaluation of the effects of social distancing measures on the reduction of the rate of virus spread in terms of reproduction number (expected number of secondary cases generated by one case);
    \item feed models to estimate the economic costs of the different interventions, as well as the impact of control extended measures on intra-EU cross border flows and traffic jams due to the epidemic; and
    \item cover all Member States in order to acquire insights."
\end{itemize}

The aim of the initiative was in line with the European Commission Recommendation to support exit strategies through mobile data and apps \citep{COM2020c}. Adding details on how mobile positioning data can contribute to epidemiological models, the \textit{Joint European Roadmap towards lifting COVID-19 containment measures}\footnote{Joint European Roadmap towards lifting COVID-19 containment measures: \url{https://ec.europa.eu/info/sites/info/files/communication_-_a_european_roadmap_to_lifting_coronavirus_containment_measures_0.pdf}} of 15 April  explains that ``[$\ldots$] mobile network operators can offer a wealth of data on mobility, social interactions [$\ldots$] Such data, if pooled and used in anonymised, aggregated format in compliance with EU data protection and privacy rules, could contribute to improve the quality of modelling and forecasting for the pandemic at EU level."  It is worth mentioning that MNO data have successfully been used to study and respond to epidemics in the recent past as \textit{e.g.} in \cite{Broadband}.

The data sharing initiative is in line with the EU ambition to become a “leading role model for a society empowered by data to make better decisions” as outlined in the Commission’s communication entitled \textit{A European strategy for data} \citep{COM2020d} released in February 2020 just before the outbreak of the pandemic. This strategy drew from a number of efforts including the High-Level Expert Group on B2G data sharing, set up by the Commission in autumn 2018 and whose members represented a broad range of interests and sectors. In its final report issued in February 2020 (\cite{alemanno2020towards}), the Expert Group called the  Commission,  the  Member  States  and  all  stakeholders  to  take  the  necessary  steps to make more private data available and increase its reuse for the common good. These previous efforts proved very timely and paved the way for the COVID-19 data sharing initiative.

 The European geographical scale of the MNOs involved in this exercise, through the processing of aggregate and anonymised data,  aims at the understanding and sharing of best practices across countries, highlighting which mobility policies are the most effective to fight COVID-19.    The process applied by the operators transforms the raw mobile data\footnote{In this paper, the term “raw data” means mobile phone data records or collection of variables that refer to individual users – not necessarily identifiable - and not to groups of users.} into aggregate and anonymised intermediate products (so called Origin-Destination Matrices, see next section for more details). These matrices are the products delivered by the MNOs to the Commission. The matrices are not “ready-to-use” indicators, but their level of granularity and their attributes have given the Commission the opportunity to derive from them indicators specifically designed to meet the needs of JRC researchers, the ECDC\footnote{ECDC: European Centre for Disease Prevention and Control. An agency of the European Union.}, policymakers and practitioners at EU and Member States level. In addition to the increased flexibility to design indicators tailored to the policymakers needs, this arrangement also gives the possibility to reduce the ‘black box’ effect thanks to the greater transparency and control over the process. This results in output indicators that are more usable by policymakers.

Even though some mobility data derived from social media\footnote{\url{https://www.google.com/covid19/mobility/}} or mobile app\footnote{\url{https://covid19.apple.com/mobility}} location data is openly available, contains rich disaggregations (\textit{e.g.} walking mobility versus driving, or categorisation of places in retail, parks, etc.) and has global coverage, the MNO data available provide a number of advantages over that location data: level of granularity, both spatial (MNO data reaches up to municipality level while app data available is limited to region or province) and temporal (some MNOs provide more than daily updates); representativity (MNO data probably better captures all the different population groups); availability of connectivity data ( \textit{i.e.}  from an origin to a destination, as opposed to just mobility levels at a location); a higher level of transparency (more detailed methodological description). This makes the MNO data a very valuable source of human mobility insights.

The unique nature of the initiative lies in its geographical scope, the number of involved Mobile Network Operators, and their relatively rapid and in many cases unconditional support offered. Thanks to continuous dialogue with the Commission, MNOs have shown concrete interest in being active and supportive, irrespective of the different levels of maturity in producing the required data; some were already collecting and processing aggregate and anonymised data to deliver similar insights to national authorities, others had to develop \textit{ad hoc} processes to be in a position to respond to the data request. Within a few months, data from  17 MNOs covering 22 EU Member States plus Norway have been transferred to the Commission on a daily basis, with an average latency of a few days, and in most cases covering historical data from February 2020. This enables the comparative analysis across countries of mobility before, during and after the release of lock-down measures.

\section{Mobility Data: Definitions, Harmonisation and Comparability across Providers}\label{sec:mobdata}

The urgency of producing useful insights and quickly supporting the response to the crisis, combined with the fact that the initiative is \textit{pro-bono}, led to the decision of sharing Origin-Destination Matrix (ODM) data already available, requiring the least possible additional developments by the MNOs. As expected, the shared ODMs follow definitions and methodologies that inevitably differ among MNOs in various aspects, such as:

\begin{itemize}
    \item \textit{Geolocation}. Movements are based on the processing of a series of positions obtained through the association to the network antenna to which users are connected, or the centroid of the cell although there are techniques to increase the geolocation accuracy  (see \cite{ricciato2020towards}) for a complete overview of different geolocation schemes)  MNOs log the communication activity (‘events’) between the user’s mobile phone and the network. These events are referred to as Call Detail Records (CDR), which include mobile phone calls, messaging, and internet data accesses, or more generally as eXtended Detail Records (XDR), which also include network signalling data.   
    \item \textit{Mobility definitions}. Several operators register a movement when a user is in a new destination area for a minimum period of time known as ’stop time’ which might vary across MNOs from 15 to 60 minutes. A few MNOs use a different approach and record a movement when users spend most of the time within a time window of  \textit{e.g.}  8 hours in an area different than the area of the previous time window or the area where they usually spend most of the time during the night. As a result, the same physical mobility pattern can be captured and described by different origin-destination movements depending on the approach adopted.
    \item \textit{Extrapolation}. Some operators extrapolate the movements counts to the total population based on their market share in the country.
    \item \textit{Spatial and temporal resolution of ODMs}. MNOs use different types of geographical areas for capturing movements, for example, administrative boundaries (such as municipalities, postcodes, and census areas) or regular geo-grids. Similarly, the time-frequency of the reported movements is heterogeneous, with time-windows ranging between one and twenty four hours.
    \item \textit{Confidentiality thresholds}. MNOs discards all movements below a given "confidentiality threshold" to reduce de-anonymisation risks. This confidentiality threshold is set in adequate proportion to the size of the adopted geographical area of reference (and its lowest population) and thus varies across MNOs.
    \item \textit{Syntactic heterogeneity}. MNOs deliver ODMs in different data formats, geographic files in different coordinate reference systems and use different languages to describe the data. These syntactic heterogeneity issues are of minor importance compared to the others heterogeneity issues described above and can easily be addressed.
    \item \textit{Presence of additional attributes}. Some MNOs provide additional information about mobile phone users such as age groups and sex as well as inbound or outbound roamers. As a matter of fact, the more the level of disaggregation or granularity, the more movements fall below the confidentiality threshold and are therefore filtered out.
\end{itemize}

The large variations of the parameters described above lead to a low harmonisation of the aggregate and anonymised data across operators. Through further aggregation and relativisation ( \textit{i.e.}  concentrating to mobility trends rather than absolute figures), it was possible to use the principle of \textit{common denominator} to derive from the ODMs a number of mobility data products that preserve a certain amount of basic shared characteristics and are therefore mostly comparable across countries. Through the process of harmonisation the ODMs received from the MNOs (and that differ in geolocation, mobility definitions, etc.) are transformed into mobility data products with a higher degree of comparability across MNOs than the original matrices.  Although the derived insights have proved extremely useful, constraints of data heterogeneity across operators could be overcome by formulating refined data requests following Trusted Smart Statistics\footnote{The Trusted Smart Statistics is a concept that describes the ongoing efforts to augment the established components of statistical systems with the elements necessary to successfully exploit the increased datafication of society. These efforts involve the working models, operational processes and practices of statistical offices.} concept as in \cite{ricciato_wirthmann_hahn_2020}, with the results of providing harmonised statistics on human mobility to more effectively feed epidemiological models.

\section{First insights - Mobility Data Products}\label{sec:prods}

 The first results of the initiative were communicated to the general public by the European Commission daily news\footnote{\url{https://ec.europa.eu/commission/presscorner/detail/en/mex_20_1359}} on 15 July 2020, by DG JRC\footnote{\url{https://ec.europa.eu/jrc/en/news/coronavirus-mobility-data-provides-insights-virus-spread-and-containment-help-inform-future}} and DG CONNECT\footnote{\url{https://ec.europa.eu/digital-single-market/en/news/coronavirus-mobility-data-provides-insights-virus-spread-and-containment-help-inform-future}}. The first results were covered by official channels, such as,  national\footnote{\url{https://www.rtp.pt/noticias/economia/operadoras-em-portugal-e-outros-18-paises-da-ue-ja-forneceram-dados-a-bruxelas_n1245101}} \footnote{ 
 \url{https://eng.belta.by/partner\_news/view/eu-study-on-mobile-phone-data-reveals-correlation-between-human-mobility-covid-19-spread-131786-2020/}} 
 and international \footnote{\url{https://www.eureporter.co/frontpage/2020/07/16/coronavirus-mobility-data-provides-insights-into-virus-spread-and-containment-to-help-inform-future-responses/}} \footnote{\url{http://www.xinhuanet.com/english/2020-07/16/c_139215507.htm}}news outlets, medical based news \footnote{\url{ https://eurohealthnet.eu/newsletter-article-hh/july-2020/}} and EUreporter \footnote{\url{ https://www.eureporter.co/frontpage/2020/07/16/coronavirus-mobility-data-provides-insights-into-virus-spread-and-containment-to-help-inform-future-responses/}}. Messages in social media were positively received.
 
 Three Mobility Data Products have been developed during the first phase of the initiative. The first two, referred to as the \textit{Common Denominator}, are the result of space-time aggregation and normalisation of the ODMs, aimed to provide further safeguard in terms of both users’ privacy and commercial sensitivity of the data, and mostly to allow comparability across countries. The third product, referred to as Mobility Functional Areas (MFAs) allows identifying areas with a high degree of inter-mobility exchange.  The three products, which are described in the following paragraphs, are being shared with authorised JRC researchers for COVID-19 related research.
 
\subsection{Mobility Indicators}
The Mobility Indicators aggregate further the MNO-provided ODMs to common spatial and temporal granularities, thus allowing for an easier comparison of mobility patterns both in time and across different European countries. The indicator uses the NUTS (Nomenclature of Territorial Units for Statistics) regions as standardised geographical reference. Specifically, it shows mobility patterns from NUTS0 (country) to NUTS3 level. For a given geographic area, the Mobility Indicator provides a historical time series of mobility according to the direction of the movements as internal, inwards, outwards and total. For more information about the ‘‘Mobility Indicators'' and their applications, we redirect the interested reader to \cite{SANTAMARIA2020104925}.

\subsection{Connectivity Matrix}
The Connectivity Matrix, similarly to the Mobility Indicators,  aggregates further the ODMs to a common space-time granularity. Unlike the Mobility Indicator, the Connectivity Matrix provides information about bilateral movements between areas. The indicator adopts a weekly time frequency (separating between week days and weekend mobility), and uses the NUTS3 regions as a geographic reference. For more information about the Connectivity Matrix and its applications, we redirect the interested reader to \cite{Iacus2020}.

\subsection{Mobility Functional Areas}
\begin{figure}[H] 
    \centering
    \includegraphics[width=0.48\textwidth]{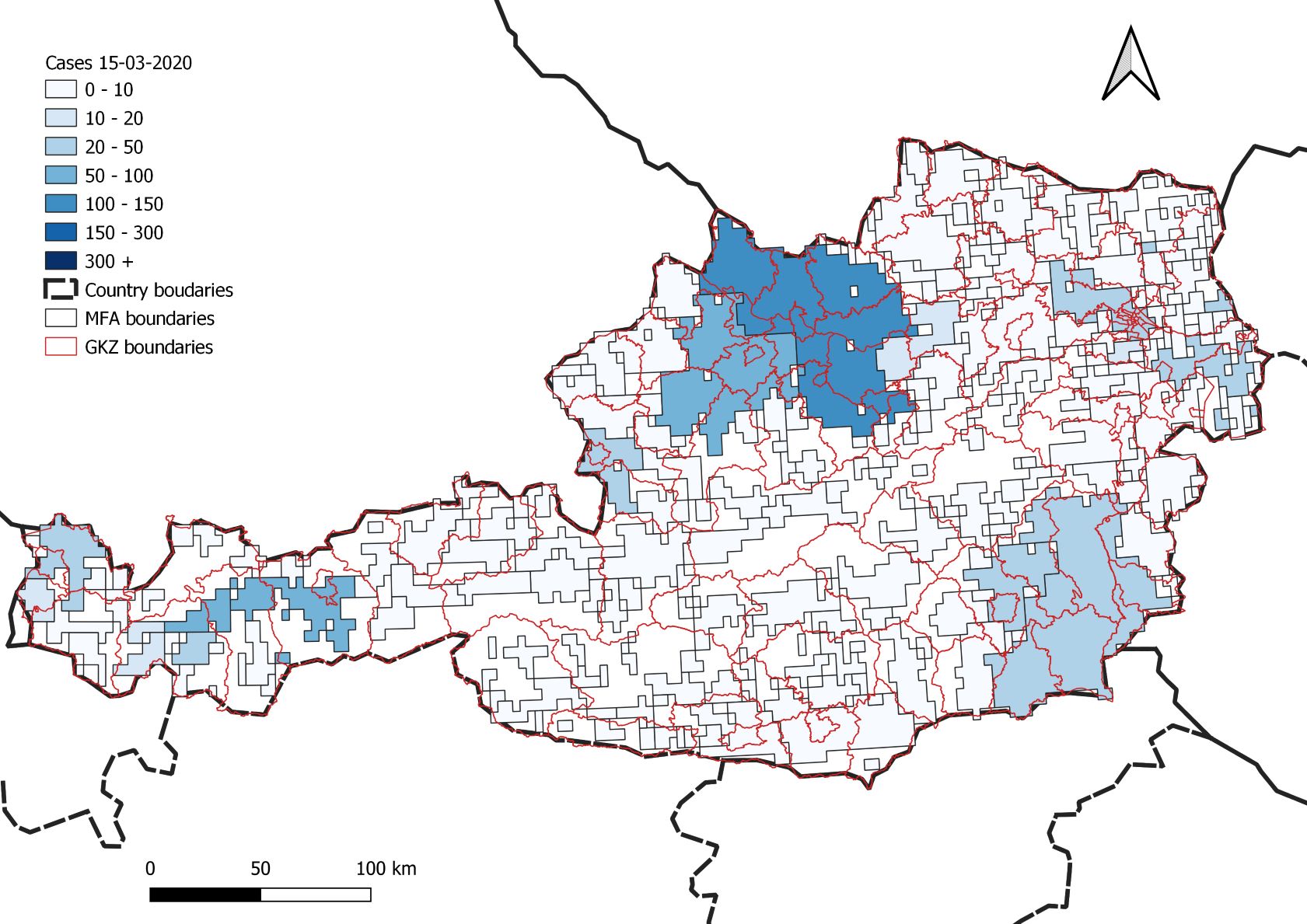}
    \includegraphics[width=0.48\textwidth]{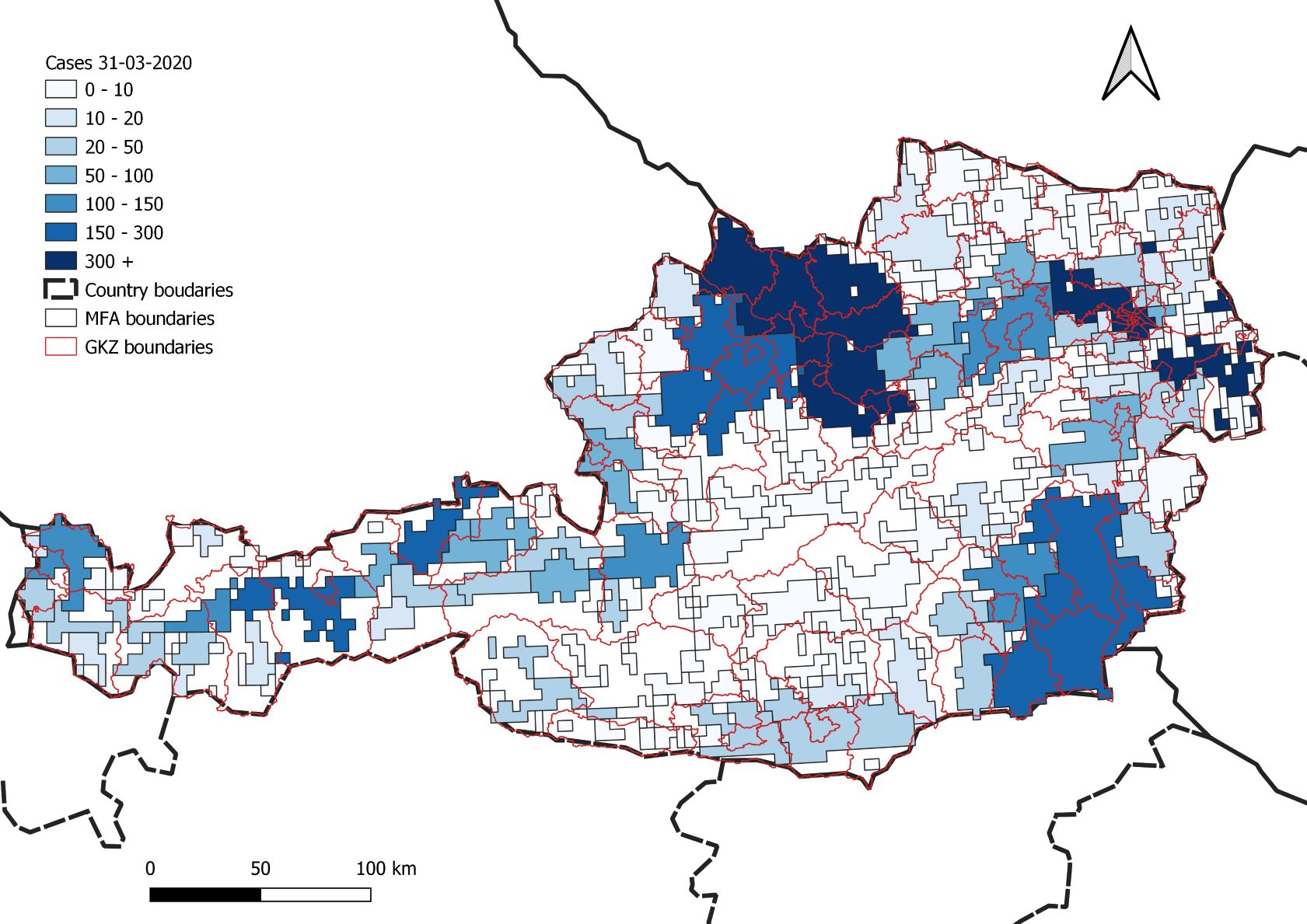}
        \includegraphics[width=0.48\textwidth]{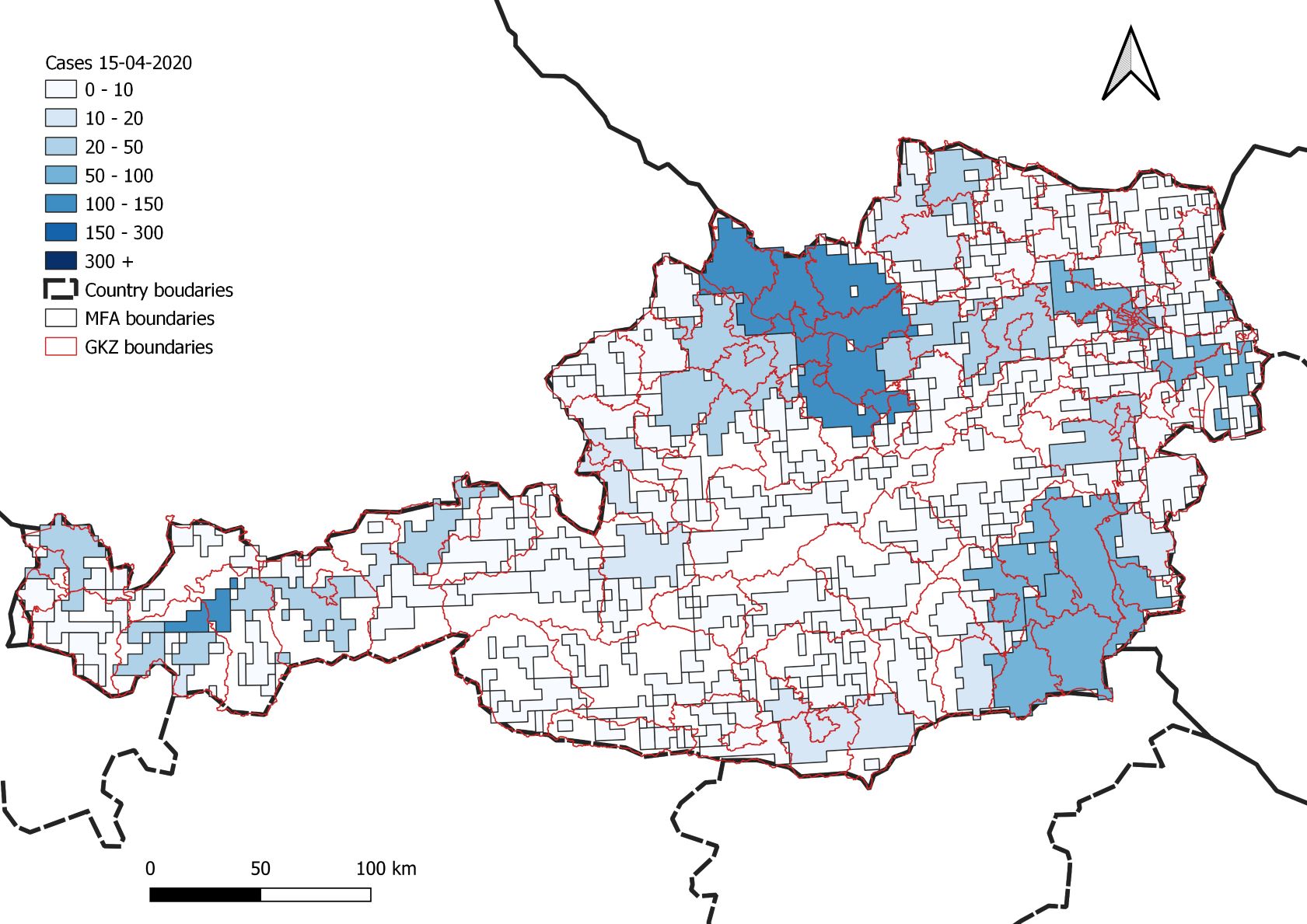}
        \includegraphics[width=0.48\textwidth]{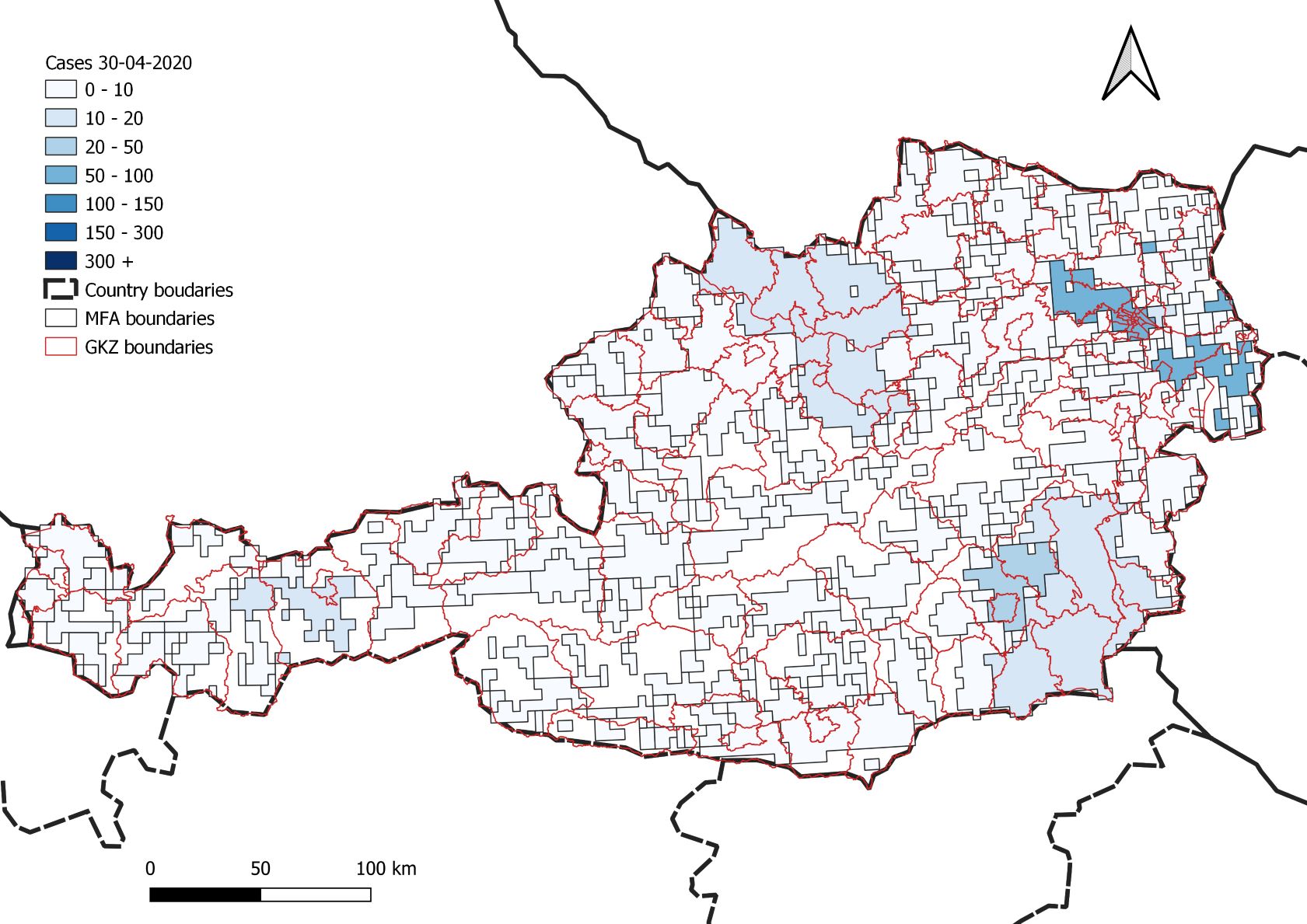}
\caption{Number of cases in previous 7 days for the dates 15 and 31 March (top), 15 and 30 April 2020 (bottom) over MFAs in Austria. The COVID-19 geographic spread seems to follow MFAs more than the  the number of political districts (GKZ) borders (\cite{iacus2021mobility}).}
    \label{fig:mfaCases}
\end{figure}

The Mobility  Functional Areas (MFAs) are data-driven geographic zones with a high degree of inter-mobility exchange. The construction of the MFAs, which is completely data-driven, starts from the ODMs at the highest spatial granularity available irrespective of administrative borders. The construction of the daily MFAs consists in identifying, through a network analysis approach, disjoint clusters of origins and destinations,  \textit{i.e.}  subsets of the ODM, that exchange a high number of movements among them. Some of the  MFAs change daily following mobility patterns. The ‘persistent’ MFAs, called simply MFAs, are obtained  through the fuzzy intersection of the daily MFAs. The rationale behind the MFAs is that the implementation of different physical distancing strategies (such as school closures or other human mobility restrictions) based on MFAs instead of administrative borders might lead to a better balance between the expected positive effect on public health and the negative socio-economic impact as discussed in \cite{TR-MFA}. As an example, it has been possible to see that the number of COVID-19 cases follow geographical patterns similar to the shapes of the MFAs. This is qualitatively shown in Figure \ref{fig:mfaCases} for Austria, and further analysed in \cite{iacus2021mobility}, suggesting that policy measures could be effectively taken on the basis of MFAs rather than on administrative borders. 


\subsection{Outreach}
\begin{figure}[H]
    \centering
    \includegraphics[width=.95\textwidth]{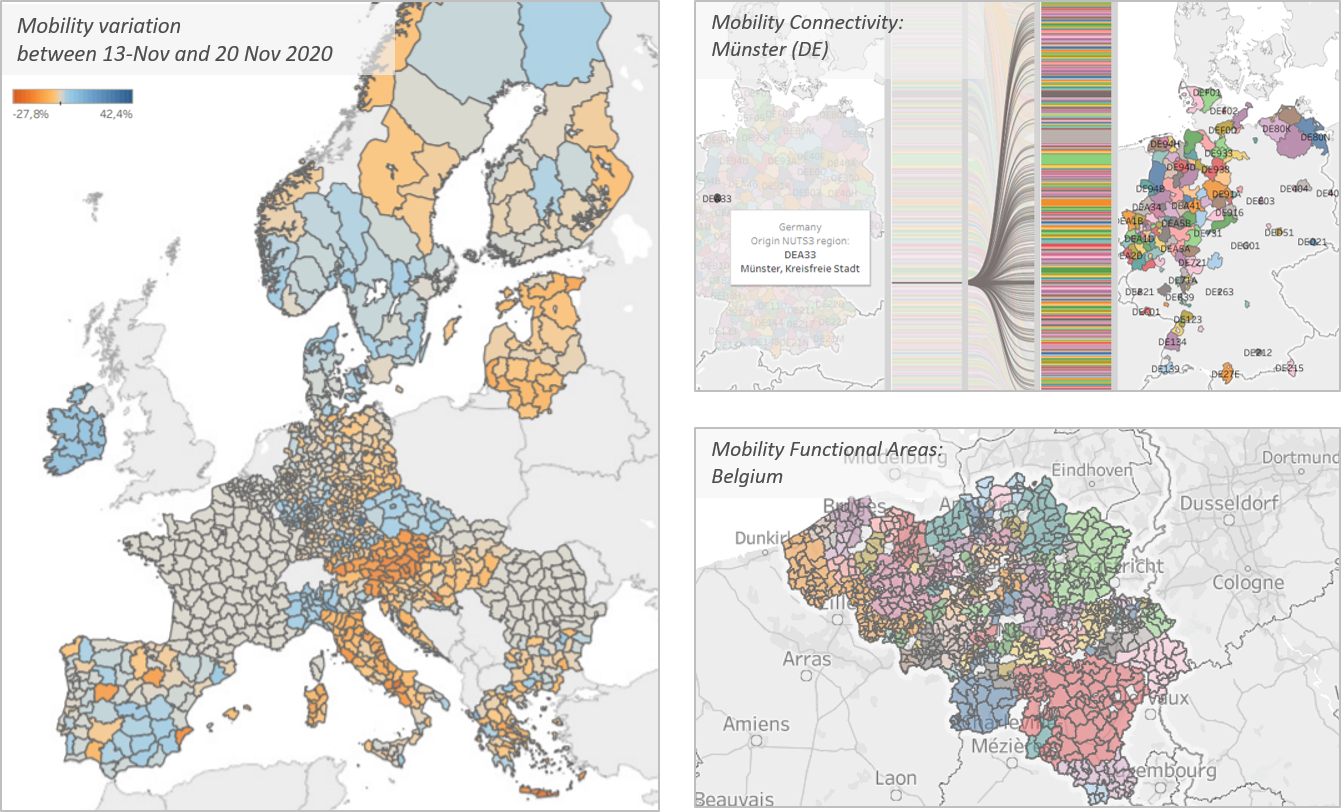}
    \caption{The Mobility Visualisation Platform allows the access and visualisation of the three products (left: mobility indicators, top right: connectivity matrices and bottom right: Mobility Functional Areas), presenting insights comparable at national, regional and NUTS3 level and combining ECDC data. Access to the platform is provided to practitioners and policymakers in the Commission, ECDC and EU Member States.}
    \label{fig:MVP}
\end{figure}

The products are currently feeding the \textit{Mobility Visualisation Platform} (Figure \ref{fig:MVP}), specifically developed to facilitate the access to the insights derived for the benefit of EU Member States, in particular through the eHealth Network\footnote{eHealth Network: \url{https://ec.europa.eu/health/ehealth/policy/network_en}}.  Moreover, the \textit{Staying safe from COVID-19 during winter} strategy adopted by the European Commission in December 2020 mentions: ``Insights into mobility patterns and role in both the disease spread and containment should ideally feed into such targeted measures. The Commission has used anonymised and aggregated mobile network operators’ data to derive mobility insights  and build tools to inform better targeted measures, in a Mobility Visualisation Platform, available to the Member States. Mobility insights are also useful in monitoring the effectiveness of measures once imposed.'' (\citeauthor{COM2020}). 

The products are currently being expanded to feed early warning mechanisms to detect anomalies in usual mobility patterns such as gatherings (\cite{Iacus2021}) and to inform scenarios for targeted COVID-19 non-pharmaceutical interventions (\cite{DeGroeve}).

\section{Challenges and Recommendations for the Future}

Because of its unprecedented nature, this B2G initiative highlighted some complex challenges that need to be addressed in order to benefit from the lessons learned. The most relevant challenges are listed below by main domain.

\begin{itemize}
\item \textit{Data Security and Integrity}.  Security and integrity of the data were primarily addressed by implementing end-to-end encryption to the data transferred from the MNOs to the JRC, and by developing a dedicated secure platform to host and process the data which is  accessible by a limited and controlled number of users. Data received by the MNOs at high spatial and temporal resolution was not allowed to exit  the secure platform. All the data processing, analysis and storage took place remotely on the Unix secure platform using open-source technologies such as Python, R and PostgreSQL. The \textit{Common Denominator} resulting from this process is represented by the Mobility Data products presented in Section~\ref{sec:prods}. 
\item \textit{Privacy, Commercial Sensititivity and Fundamental Rights}. Data privacy, risk of re-identification of groups of individuals and ethical aspects related to the use of the data needed to be carefully addressed. Although the data shared by the MNOs (ODMs) contains only anonymised and aggregate data, in compliance with the EDPB guidelines (\cite{edpb2020}) the JRC carried out a so-called ``Reasonability Test'' upon the reception of preliminary data samples from MNOs. The objective of the test is twofold: to actively verify that the data specification in terms of origin destination aggregate data were respected and to assess whether or not the risk of re-identification of the individuals was reasonably low. Following the recommendations by the European Data Protection Supervisor (EDPS), the JRC Data Protection Coordinator and the GSMA COVID-19 Privacy Guidelines (\cite{GSMA2020}), and in order to respect fundamental rights, avoid discrimination as well as respect of legitimate business interests of operators, the JRC put in place measures such as: \textit{i)} definition of conditions of non disclosure and use of the data products only in well identified COVID-19 related fields (as set out in the Letter of Intent (see Section \ref{Intro}); \textit{ii)} limited and controlled access not only to the original MNO data but also to the derived products; and \textit{iii)} adoption of a data retention horizon.
\item \textit{Communication and transparency}. Communication aspects were duly analysed in order to appropriately convey the message that the initiative has dealt exclusively with anonymised and aggregate data, ultimately avoiding reputational damage both for the MNOs and the Commission as well as political backlash. This required consultation with MNOs and the GSMA prior the publication of communication outlets as well as scientific results to the public.
\item \textit{Data heterogeneity}. Because of the need to react quickly to an emergency situation, the initiative has been based mostly on data already available at the MNOs. Yet, the JRC had to cope with a high degree of heterogeneity of the data as introduced in Section \ref{sec:mobdata}. This implied substantial downstream technical efforts by the JRC to harmonise the data to the greatest possible extent, and to find a \textit{common denominator} across operators resulting in lowered information content but guaranteeing data comparability.
\end{itemize}


This unprecedented initiative demonstrated the importance of an inter-disciplinary approach, gathering together lawyers, epidemiologists, telecommunication engineers, data scientists, software developers, communication specialists and policymakers to face all the different relevant aspects (legal, scientific, technical, communications etc.). Well established data stewardship skills are required for successful B2G initiatives within the private, policy and scientific sectors. 

Beside all these considerations and recommendations, a question still remains unanswered: what else can we do to strengthen the preparedness for future pandemics? It has been clear from this exercise that a fast and systematic response is necessary to face future crisis situations. Updated protocols and guidelines need to be already in place at the time of the crisis to avoid potential deadlocks. In this vein, the recent Proposal for a Regulation on European data governance (Data Governance Act, \citeauthor{COM2020b}) aims to foster the availability of data for use by increasing trust in data intermediaries and by strengthening data-sharing mechanisms across the EU. 

The followings are a set of possible recommendations for future similar initiatives aiming at harnessing the potential of mobility insights for policy:
    \begin{itemize}
        \item Establish a \textit{Multi-disciplinary Working Group of Experts} (telecommunication engineers, data scientist, epidemiologists, lawyers, data protection and ethics experts, IT security experts, modellers and statisticians) to draft data specification on standardised mobility data addressing scientific challenges and to develop data security and protection protocols and avoid unilateral points of view \footnote{A few of such experiments already exist. One of this is the Social Science One (\url{https://socialscience.one/}) initiative, which is a partnership between academia and industry. The expert group is nominated and funded by independent public foundations and works closely with experts within the industry to create a framework that allows access to the industry data preserving both the privacy of the users and the commercial value of the companies involved.   The composition of the team of experts and the length of support to the initiative is under the control of the foundations. Another example is the is the Big Data for Migration Alliance (\url{https://data4migration.org/)}. }. Such interdisciplinary working group should also suggest the right balance between privacy-compliance and level-of-detail starting from the raw mobile positioning data; this may serve the MNOs to find a common standard for the ODMs, drastically reducing the heterogeneity. At the same time, the standards should not be too prescriptive to keep MNOs from innovating. Moreover, the working group should develop additional Mobility Data Products tailored to their different applications (epidemiological modelling, forecasting of the contagion on different scales, mobility-impact assessment, drop in connectivity and tourism, etc.).
        \item Establish an \textit{Ethic Committee} with the mission of considering all ethical aspects and implication of the initiative and to make sure that it complies in any of its part with fundamental rights. The Ethics Committee should take into consideration the culturally shared privacy norms of different countries involved in the initiative.
        \item Establish a \textit{Communication Team} where all the players involved in the initiative are represented; the team should be responsible to define an appropriate communication strategy for the initiative and its scientific results to the general public. The team should consider the high sensitivity of the topic, safeguarding the reputation of both the involved Institutions and the private partners, while preventing misconceptions and misunderstandings about the use of mobile phone data.
        \item Draft a \textit{Risk Assessment} pointing out all possible risks in the various stages of the initiative and defining their likelihood, their possible impact on the success of the project and their mitigation strategies. The risk assessment allows anticipating most of the possible faux pas, carelessness or mistakes that might jeopardise the success of similar initiatives. The risk assessment should address security, privacy and ethical threats, with appropriate risk management plan.
        \item Establish and maintain a \textit{Research Network of Collaboration} made of data providers, data processors and data users. Such a network will facilitate the dialog between the parties, ensuring the continuous improvement of procedures and the enhancement of the results.
\end{itemize}

In terms of highest priority, the \textit{ex-ant}e definition of a common standard for the raw data from the MNOs is indeed the first step, since it requires a long time to be drafted and an even longer time to be implemented. Moreover, its adoption would ensure direct comparability across both regions and mobile data sources, while avoiding the time-and-resource demand in harmonisation at the data processor side (see the heterogeneous characteristics of the ODMs explained in Section~\ref{sec:mobdata}), which not only requires an unavoidable loss of space-time granularity, but very often leads to sub-optimal indicators. 

Efforts should also aim at making insights and derived products publicly available to the research community in order to make use of its full potential and ensure reproducibility of research outcomes.

\section*{Conclusions and Next Steps}

Based on the results obtained, this initiative can potentially help simplify the systematic use of data from the private sector in the framework of policy support. The global scale and spread of the COVID-19 pandemic highlight the need for a more harmonised or coordinated approach across countries \citep{Olivereabc0764}. By collecting mobility data across various EU member states, this initiative is trying to address COVID-19 crisis response in a more holistic way. The initiative provides a concrete example on setting up bilateral channels between private and public sectors in understanding human mobility and providing support in addressing societal issues. Some of the procedures developed during the project should be consolidated in order to speed up future B2G processes. The success of the initiative demonstrates how actors at the interface between the private sector, the scientific community and the policy side can play a key  intermediary role in ensuring that data driven insights are used responsibly to effectively respond to pressing policy questions.

\paragraph{Acknowledgments}
 The authors acknowledge the support of European MNOs (among which 3 Group - part of CK Hutchison, A1 Telekom Austria Group, Altice Portugal, Deutsche Telekom, Orange, Proximus, TIM Telecom Italia, Tele2, Telefonica, Telenor, Telia Company and Vodafone) in providing access to aggregate and anonymised data, an invaluable contribution to the initiative. The authors would also like to acknowledge the GSMA\footnote{GSMA is the GSM Association of Mobile Network Operators.}, colleagues from Eurostat\footnote{Eurostat is the Statistical Office of the European Union.} and ECDC  for their input in drafting the data request. 
Finally, the authors would also like to acknowledge the support from JRC colleagues, and in particular the E3 Unit, for setting up a secure environment to host and process of the data provided by MNOs, as well as the E6 Unit (the “Dynamic Data Hub team”) for their valuable support in setting up the data lake.

\paragraph{Competing interests}
None

\paragraph{Ethical standards}
The research meets all ethical guidelines.

\paragraph{Author contributions}
 All authors equally contributed to the work. All authors approved the final submitted draft. 

\paragraph{Funding statement} This work received no specific grant from any funding agency, commercial or not-for-profit sectors.

\paragraph{Data Availability Statement} No datasets were processed to generate research results presented in the current study.

\bibliographystyle{apalike}
\bibliography{References}

\end{document}